\documentclass[pra,aps,twocolumn,superscriptaddress]{revtex4-1}
\usepackage{latexsym,amssymb,amsfonts,amsmath,amstext,graphicx,bbm,bm,relsize,dsfont,theorem,times}
\usepackage[dvipsnames]{xcolor}
\usepackage[colorlinks=true,linkcolor=blue,urlcolor=blue,citecolor=blue]{hyperref}
\newcommand{\tr}{\mbox{tr}}

\def\tr{\mbox{tr}}
\def\bra#1{\langle{#1}|}
\def\ket#1{|{#1}\rangle}

{\catcode`\|=\active
  \gdef\Braket#1{\begingroup
\mathcode`\|32768\let|\BraVert\left<{#1}\right>\endgroup}}
\def\BraVert{\egroup\,\mid\,\bgroup}

\newtheorem{definition}{Definition}

\newtheorem{observation}[definition]{Observation}

\newcommand{\Var}{\textrm{Var}}

\newcommand{\mean}[1]{\langle #1 \rangle}

\newcommand{\Ham}{{H}}

\newcommand{\Id}{\mathds{1}}

\allowdisplaybreaks

\begin{document}

\title{Quantum work statistics and resource theories: bridging the gap through R{\'e}nyi divergences}

\author{G. Guarnieri}
\email{gguarnieri88@gmail.com}
\affiliation{Department of Optics, Palack\'{y} University, 17. Listopadu 1192/12, 771 46 Olomouc, Czech Republic}

\author{N. H. Y. Ng}
\affiliation{Dahlem Center for Complex Quantum Systems, Freie Universit\"{a}t Berlin, 14195 Berlin, Germany}

\author{K. Modi}
\affiliation{School of Physics and Astronomy, Monash University, Victoria 3800, Australia}

\author{J.~Eisert}
\affiliation{Dahlem Center for Complex Quantum Systems, Freie Universit\"{a}t Berlin, 14195 Berlin, Germany}

\author{M. Paternostro}
\affiliation{Centre for Theoretical Atomic, Molecular and Optical Physics, Queen's University Belfast, Belfast BT7 1NN, United Kingdom}

\author{J. Goold}
\affiliation{School of Physics, Trinity College Dublin, Dublin 2, Ireland}
\date{\today}

\begin{abstract}
{The work performed on or extracted from a non-autonomous quantum system described by means of a two-point projective-measurement approach takes the form of a stochastic variable.  We 
show that the cumulant generating function of work can be recast in the form of quantum R\'enyi-$\alpha$ divergences, and by exploiting convexity of this cumulant generating function, derive
a single-parameter family of bounds for the first moment of work. Higher order moments of work can also be obtained from this result. In this way, we establish a link between quantum work statistics in stochastic approaches
on the one hand
and resource theories for quantum thermodynamics on the other hand,
a theory in which R\'enyi-$\alpha$ divergences take a central role.
To explore this connection further, we consider an extended framework involving a control switch and an auxiliary battery, which is instrumental to reconstruct the work statistics of the system. 
We compare and discuss our bounds on the work distribution to 
findings on deterministic work studied in resource theoretic settings. 
}

\end{abstract} 
\maketitle

\makeatletter
 
Fundamental out-of-equilibrium fluctuation theorems have been formulated to characterize the full non-linear response of both classical and quantum systems, to the action of a time-dependent external perturbation~ \cite{jarzynski2011equalities,seifert2012stochastic,esposito2009nonequilibrium,campisi2011colloquium,hanggi2015other}. Such theorems can be seen as refined statements of the second law of thermodynamics suitable for application at the nano-scale. As such, they play an important role in the characterization of quantum and classical thermodynamic processes and thermal machines~\cite{Millen:16,Goold:16,Vinjanampathy:15, Binder15a}. 
In such a framework, the statistics of stochastic thermodynamic variables can be gathered through two-time projective-measurement protocols, where the fluctuating work done by or on a system driven out of equilibrium or the heat that it exchanges with an environment are defined in terms of the difference of energy eigenvalues observed at the start and the end of the dynamics~\cite{talkner2007fluctuation,talkner2009jstat}. This approach is experimentally viable~\cite{batalhao2014experimental,an2015experimental}, has been useful for the characterization of non-equilibrium features of quenched many-body systems~\cite{silva2008statistics,gambassi2012large,mascarenhas2014work,sindona2014statistics,fusco2014prx} and there is strong evidence that it has a physically meaningful semi-classical limit~\cite{jarzynski2015quantum,zhu2016quantum,wang2017understanding, garcia2017quantum,garcia2017semiclassical}. 

A complementing and in many ways radically 
different formal approach to the stochastic thermodynamics of quantum systems is based on resource theories. 
These are mathematical frameworks that specify restrictions to the transformations that
can be performed on a quantum system to identify sets of \emph{free states} that can be prepared under such constraints. Any state $\rho$ that is not free can then be consumed to create final states which are also not free,
making $\rho$ a useful resource~\cite{brandao2015reversible}. Initially applied to entanglement and coherence~\cite{horodecki2009quantum}, resource-theoretical approaches have also been applied to study thermodynamics of quantum processes~\cite{horodecki2013resource, brandao2013resource, brandao2015second, faist2015minimal, gour2015resource,Limits}, providing insights on the inter-convertibility of finite resource states and on extractable work \cite{Aberg2013NatCom, EgloffNJP2015,Work} from a general, axiomatic perspective that does not rely on specific details of a particular system.
Despite some success in capturing fundamental aspects of the interplay between non-equilibrium thermodynamics and quantum dynamics, it is not yet ascertained whether resource theories are mutually compatible with fluctuation theorem settings, and if their predictive powers are equivalent. Moreover, their predictive statements are rarely phrased in a directly comparable manner, the only exception insofar being in the thermodynamic limit. In such case the fluctuations become negligible, and the optimal amount of work extracted is found to be given as a function of the non-equilibrium free energy of the system \cite{horodecki2013resource}.

In this work, we make a step forward towards bridging the gap between the resource theoretic approach to thermodynamics of quantum systems and the quantum stochastic one by presenting a situation in which these two frameworks may be directly compared.
In order to provide the foundations of our approach, we start by considering closed unitary dynamics for the system and, after showing that the cumulant generating function can always be recast in terms of quantum R\'enyi-$\alpha$ divergences, we derive a family of single parameter bounds on the average work. Moreover, the relation between the cumulant generating function and the R\'enyi divergences allows to relate higher moments of the work distribution to higher derivatives of the latter. The approach take is further motivated by recent insights into
modified versions of the Crooks relation in the context of single-shot work extraction \cite{Halpern} and
one-shot dissipated work from Renyi divergences \cite{Halpern2}.
We compare our result to findings from a resource theory perspective, where extractable work can also be phrased in terms of R\'enyi divergences, and discuss the similarities and distinctions of these two results.

We then consider an open quantum system scenario, where the system of interest interacts with a thermal bath and is attached to both an auxiliary quantum battery and a control switch (the latter describing the action of an external driving work protocol) so as to meet the usual conditions invoked in resource theory-based approaches. As a general prescription, we perform the two-point measurement protocol on the auxiliary battery rather than on the system, thus preserving any initial coherence in the latter. Work can then be consistently defined in terms of the energy difference in the battery.
This allows us to show that, when the system is initially prepared in a thermal state (which is anyway implicitly assumed in every explicit quantum stochastic approach \cite{jarzynski2015quantum, Campisi2010RMP}) the cumulant generating function can still be cast in terms of quantum R\'enyi-$\alpha$ divergences. Analogue considerations and results as done in the first part of the work for closed systems are thus recovered in this extended scenario. In particular, this provides strong (or significant) evidence of the equivalence. Moreover, our results allow to attach a clear physical interpretation to the $\alpha$-R\'enyi divergences by linking them to statistical quantities that are experimentally accessible.

{\it Stochastic approach.} Consider an isolated quantum system -- initially prepared in an equilibrium state at inverse temperature $\beta>0$ -- subjected to an external force that changes a {\it work} parameter $\lambda_t$ in time according to a generic finite-time protocol. The latter includes, at the initial time $t=0$ and final time $t=\tau$, projective measurements of the energy of the system, which results in the values $E^{\lambda_0}_n$ and $E_m^{\lambda_\tau}$. Here, $n$ and $m$ labels the respective energy levels of the initial and final Hamiltonian ${\cal H}(\lambda_0)$, ${\cal H}(\lambda_\tau)$ of the system. Thermal and quantum randomness render the measured energy difference $E_m^{\lambda_\tau} - E_n^{\lambda_0}$, which can be interpreted as the work done on the system through the protocol, a stochastic variable whose values are provided by the trajectory-ensemble distribution~\cite{talkner2007fluctuation}
\begin{equation}\label{eq:probdistrib}
p_\tau(W) = \sum_{n,m} P_\tau\left[ E_m^{\lambda_\tau} , E_n^{\lambda_0} \right] \delta\left[W -(E_m^{\lambda_\tau} - E_n^{\lambda_0})\right].
\end{equation} 
Here, $P_\tau[ E_m^{\lambda_\tau} , E_n^{\lambda_0}]$ is the joint probability density
that the two-time energy measurements results in the values $E_m^{\lambda_\tau}$ and $E_n^{\lambda_0}$.
The Fourier transform of $p_\tau$ gives a generating function 
$\Theta(\eta ,\tau) := \langle  e^{i \eta W} \rangle_\tau$
which, by derivation over the counting field parameter $\eta$, gives the $n^{\rm th}$-order moment of work.
Another informative quantity that we shall consider in this work is the cumulant generating function 
\begin{equation}\label{eq:LaplCumGen}
\Phi(\eta, \tau) := \ln \langle  e^{- \eta W} \rangle_\tau = \ln \int dW\, p_\tau(W) e^{- \eta W}.
\end{equation}
The quantity $(-1)^n{\partial^n_\eta}\Phi(\eta,\tau)|_{\eta=0}$, gives us the cumulants of work. 
Using the H\"older inequality, it is possible to demonstrate the convexity of $\Phi$ with respect to the 
first argument \cite{touchette2009large}. 
This property can be equivalently stated as~\cite{rockafellar2015convex}
\begin{equation}\label{Convexity}
\Phi(\eta,\tau) \geq \eta {\partial_\eta}\Phi(\eta, \tau)|_{\eta=0}, 
\end{equation}
and, as $\mean{W}_\tau = -{\partial_\eta}\Phi(\eta,\tau)|_{\eta=0}$, we immediately obtain a single-parameter family of lower bounds for the mean work,
\begin{equation}
\label{eq:bound}
\beta \mean{W}_\tau  \geq -\frac{\beta}{\eta}\Phi(\eta, \tau) ,\quad \quad \eta > 0.
\end{equation}
A similar set of one-parameter bounds was recently derived in the context of Landauer erasure~\cite{guarnieri2017full}. For negative values $\eta <0$, a family of upper bounds $\beta \mean{W}_\tau \leq \beta \Phi(\eta, \tau)/|\eta|$ is obtained instead.

{\it Connecting the bounds to R\'{e}nyi divergences.} It is well known in the field of full counting statistics~\cite{esposito2009nonequilibrium} that, for an initial Gibbs state of the bath, the cumulant generating function can be recast as
\begin{equation}\label{eq:CumGenFCS}
\Phi(\eta, \tau) = \ln \mathrm{Tr}\left[\rho_S(\eta, \tau)\right]
\end{equation}
with $\rho_S(\eta, \tau) = {U}_{\eta/2}(\tau) \rho_S(0) {U}^{\dagger}_{-\eta/2}(\tau)$, with the operator ${U}_{\eta}(\tau) := e^{-\eta\Ham(\lambda_\tau)} {U}(\tau)  e^{\eta\Ham(\lambda_0)}$ and $ U(\tau)$ being the time-evolution operators of the system at time $\tau$. Starting from this expression, the following identity can be derived (see Appendix \ref{AppA} for details)
\begin{observation}[Cumulant generating function] The cumulant generating function for the moments of work
is given by
\begin{equation}
\label{eq:central}
\Phi(\eta,\tau)=-\frac{\eta}{\beta}S_{1-\frac{\eta}{\beta}}\left(\rho_S(\tau)\, ||\,\mathcal{G}_S(\lambda_\tau)\right) - \eta \Delta F,
\end{equation}
where
$\mathcal{G}_S(\lambda_t) := Z(\lambda_t)^{-1} e^{-\beta\Ham(\lambda_t)}$ denotes the canonical Gibbs state at time $t$ and $\Delta F = F(\lambda_\tau) - F(\lambda_0)$ is the free energy difference between canonical Gibbs states at 
the initial and final points, with the free energy of Gibbs states at time $ t $ being $F(\lambda_t)=-\beta^{-1}\ln Z(\lambda_t)$ and $Z(\lambda_t)$ the partition function of ${\cal H}(\lambda_t)$.
\end{observation}
In this expression, the quantum $\alpha$-R\'enyi divergences are defined as
\begin{equation}\label{eq:RelationRenyi1}
S_{\alpha}(\rho||\sigma) := \frac{1}{\alpha - 1} \ln\mathrm{Tr}\left[\rho^{\alpha}\sigma^{1-\alpha}\right],\quad \alpha \in (0,1)\cup (1,+\infty),
\end{equation}
with $\rho$ and $\sigma$ being two generic density matrices~\cite{wei2017relations}.
The R\'enyi divergence of order $ \alpha =1 $ reduces to the familiar quantum relative entropy, i.e., $ \displaystyle\lim_{\alpha\to 1} S_\alpha (\rho\|\sigma) = D(\rho\|\sigma) = \tr (\rho\ln\rho - \rho\ln\sigma)$. Eq.~\eqref{eq:RelationRenyi1} has recently gained much attention due to its role in resource-theoretical formulations of thermodynamics~\cite{brandao2015second} and the central role that it plays in the quantification of the irreversible entropy production resulting from non-equilibrium processes~\cite{santos2017prl,wei2017relations}. 
Combining  Eq.~\eqref{eq:central} with Eq.~\eqref{eq:bound}, one obtains an inequality on the irreversible entropy $\langle S_{\text{irr}} \rangle :=\beta(\langle W\rangle-\Delta F)$ ~\cite{Donald1987,deffner2010generalized},
\begin{equation} \label{eq:mainres1}
\langle S_{\text{irr}} \rangle \geq S_{1-\frac{\eta}{\beta}}\left(\rho_S(\tau)\, ||\, \mathcal{G}_S(\lambda_\tau)\right).
\end{equation} 
It is important to stress such a relation stems just from the convexity property of the cumulant generating function of the work distribution and Eq.~\eqref{eq:central}. However, 
noting Refs.\ \cite{BreuerPRA2003,ParrondoNJP2009,DeffnerLutzPRL2011} 
and accepting that the condition on R\'enyi divergences is the tightest for $\alpha=1$, 
$\langle S_{\text{irr}} \rangle = \lim_{\eta\to 0^+} S_{1-\frac{\eta}{\beta}}\left(\rho_S(\tau)\, ||\, \mathcal{G}_S(\lambda_\tau)\right)$,
we arrive at the following statement.

\begin{observation}[Stochastic irreversible entropy] The irreversible entropy
in the stochastic approach
is 
\begin{align}\label{eq:equality_Sirr}
\langle S_{\text{irr}} \rangle
&= D\left(\rho_S(\tau)\, ||\, \mathcal{G}_S(\lambda_\tau)\right).
\end{align}
\end{observation}

Furthermore, Eq.~\eqref{eq:central} can be used  to relate higher moments of the work distribution and higher derivatives of the R\'enyi-$\alpha$ divergence. In particular, the second cumulant $ {\rm Var}(W) $ is 
\begin{align}
\Var (W) &= \frac{2}{\beta^2} \frac{\partial S_\alpha (\rho_S(\tau)\| \mathcal{G}_S(\lambda_\tau) )}{\partial \alpha}\bigg|_{\alpha=1}\label{varW}\\
&= \frac{1}{\beta^2} V (\rho_S(\tau)\| \mathcal{G}_S(\lambda_\tau)),\label{eq:relentvar}
\end{align}
where $ V(\rho\|\sigma):= \tr (\rho(\log\rho-\log\sigma)^2) - (\tr(\rho(\log\rho-\log\sigma)))^2 $ is the \emph{relative entropy variance} \cite{tomamichel2013hierarchy,lin2015investigating}.
Finally, the fact that this cumulant generating function can be expressed in terms of R{\'e}nyi divergences allows 
us also to bound the fluctuations of work. By using a simple Chernoff bound, we have in fact that, for any $ k>0 $,
\begin{equation}\label{eq:Chernoff}
\Pr \left[W \geq \langle W\rangle + k\sigma_W \right] \leq \frac{1}{k^2},
\end{equation}
where $\sigma_W = \sqrt{\Var(W)}$ is the standard deviation for the distribution of $W$.
In a single instance of this thermodynamic process, $W$ may exhibit arbitrary fluctuations. However, if one considers multiple identical processes and determines the overall accumulated work, then we know via central limit theorem
that the total work assumes a sharply peaked normal distribution. More concretely, consider the amount of work $ W_n $ used to perform the stochastic process described above, for $ n $ identical copies of the initial system. In this picture, one can define the $\varepsilon$-deterministic work $W_n^\varepsilon$ to be the amount of work such that $W_n\leq W_n^\varepsilon$, except with some failure probability $\varepsilon$. The final state will be $\rho_S(\tau)^{\otimes n}$, and furthermore note that both $D(\rho_S(\tau)\| \mathcal{G}_S(\lambda_\tau))$ and $V(\rho_S(\tau)\| \mathcal{G}_S(\lambda_\tau))$ are additive under tensor product.
	Thus, for any $ \varepsilon >0 $, by substituting $k^{-2}=\varepsilon$, we conclude from Eq.~\eqref{eq:Chernoff} the following.
\begin{observation}[Stochastic $\varepsilon$-deterministic work] The work obtained in a
setting failing with probability $\varepsilon>0$ in the multi-copy stochastic approach is given by
\begin{align}\label{prob}
W_n^\varepsilon =& n \left[\mean{W} + \sqrt{\frac{\Var(W)}{\varepsilon n}}\right],
\end{align}
with $\mean{W}=\beta^{-1}[\langle S_{\text{irr}} \rangle+\Delta F]$ where $\langle S_{\text{irr}} \rangle$ was derived in Eq.~\eqref{eq:equality_Sirr}, and $\Var(W)$ in Eq.~\eqref{eq:relentvar}.
\end{observation} 

{\it General scenario and connection to resource theory.} All the results insofar have been obtained considering a closed quantum system subject to an external driving work protocol~\cite{Campisi2010RMP}.
The present aim is to demonstrate that a fundamental relation like Eqs.~\eqref{eq:central} and \eqref{eq:mainres1} can be retrieved in a different scenario closer to the typcial scenario in resource theories. 
For this reason, alongside the system of interest $S$, let us consider a control switch $C$, modeling the action of an external driving protocol, and a battery $B$ operating as a storage system for work, such that the total Hamiltonian 
is given by
\begin{equation}\label{eq:totalH}
\Ham_{S,B,C} = \Ham_S(\lambda_0)\otimes\Pi_{\lambda_0,C} + \Ham_S(\lambda_\tau)\otimes\Pi_{\lambda_\tau,C} + \Ham_B,
\end{equation}
where $\Pi_{\lambda,C}=\ket{\lambda}\bra{\lambda}_C$. 
The free Hamiltonian $H_B=X$ of the battery is taken to be a Hamiltonian 
given by the position operator.

Next, we specify the class of interactions allowed to take place between the systems $S$, $C$ and $B$.
Here, we consider operations that satisfy the following constraints:
{\bf (1)} {\it Unitarity of the dynamics of the whole compound, governed by some ${\cal U}$}. 
{\bf (2)} {\it Energy-preserving nature}, i.e., $\left[{\cal U}, \Ham_{S,B,C} \right] =0$, with $\Ham_{S,C,B}$ denoting the Hamiltonian of the overall system.
{\bf (3)} {\it Invariance under displacements of the battery}, i.e., $\left[ \Delta_B, {\cal U} \right] =0$, with $\Delta_B$ being the Weyl displacement operator shifting positions
\cite{cohen2012weyl}.
This set of operations closely resemble the set of thermal operations described in a resource theory setting \cite{alhambra2016fluctuating}, being a special case that in Eq.~\eqref{eq:totalH} there is no additional thermal bath, but instead the system itself is initialized in a Gibbs state. The inclusion of both the control switch and the battery are necessary in order to model an explicit time-dependent external work protocol into a time-independent, energy-preserving transformation \cite{Aberg2013NatCom,brandao2013resource,Aberg2016arXiv}.
While condition {\bf (1)} is comes simply from quantum mechanics, request {\bf (2)} is equivalent to asking that the Gibbs state of the global system is preserved. Finally, constraint {\bf (3)} ensures that the battery acts only as a system that stores/provides work, instead of acting as an additional resource for coherence, or as a entropy sink. Work is defined as energy difference on the battery \cite{Skrzypczyk2014NatCom},
\begin{equation}\label{eq:workDef}
W := -\left( E_n^B - E_m^B \right),
\end{equation}
where this is again a fluctuating work variable. 
The crucial difference brought along by Eq.~\eqref{eq:workDef} lies in the fact that the statistics of work is reconstructed by performing the two-projective-measurement scheme on the battery, rather than on the system. 

As a final constraint, similar to that of Ref.\ \cite{alhambra2016fluctuating}, we require that the unitary ${\cal U}(\tau)$ acting on the systems $S,B,C$, perfectly produces the desired change on the system Hamiltonian from $H_S(\lambda_0)$ to $H_S(\lambda_\tau)$. This means that, if the initial state of the control switch is taken as $\rho_C(0) = \Pi_{\lambda_0,C}$, we want to have
\begin{equation}\label{eq:globalUS,B,C}
{\cal U}(\tau) \left( \rho_{S,B}(0)\otimes\Pi_{\lambda_0,C} \right) {\cal U}^{\dagger}(\tau) = \rho_{S,B} (\tau)\otimes\Pi_{\lambda_\tau,C}.
\end{equation}
To satisfy this, we require that ${\cal U}(\tau) = {\cal U}_{S,B,1}(\tau)\otimes\ket{\lambda_\tau}\bra{\lambda_0}_C+ {\cal U}_{S,B,2}(\tau)\otimes\ket{\lambda_0}\bra{\lambda_\tau}_C,$ where ${\cal U}_{S,B,(1,2)}(\tau)$ are generic unitary transformations on the joint system $S,B$. This ensures that when $\rho_C(0)=\Pi_{\lambda_0,C}$, then 
$\mathcal{U}(\tau)$ effectively induces a unitary transformation on system $S,B$,
\begin{equation}\label{eq:redCPTP}
\rho_{S,B}(\tau)= {\cal U}_{S,B,1}(\tau)\rho_{S,B}(0) {\cal U}^{\dagger}_{S,B,1}(\tau).
\end{equation}
Condition {\bf{(2)}} expressing the energy conservation of the global system $S,B,C$ implies that 
\begin{equation}\label{eq:ConditiononU1}
\mathcal{U}_{S,B,1}(\tau) \left( \Ham_S(\lambda_0)  +\Ham_B \right)  \mathcal{U}^{\dagger}_{S,B,1}(\tau) = \Ham_S(\lambda_\tau) + \Ham_B.
\end{equation}
From this, we see that ${\cal U}_{S,B,1}(\tau)$ does not necessarily preserve the energy of $S,B$, and whatever energy difference incurred on $S,B$ is stored in the state of the switch $C$.

We demonstrate that a relation akin to Eq.~\eqref{eq:central} can be derived also in this extended scenario.
Let us consider the initial system to be prepared in $\rho_S(0) = \mathcal{G}_S(\lambda_0)$, i.e. an equilibrium Gibbs at inverse temperature $\beta$ relative to the initial Hamiltonian $\Ham_S(\lambda_0)$. Furthermore, let the initial state of the battery is a pure state $\rho_B(0)$, in a Gaussian state that well approximates a
state with definite position. Keeping in mind that after the unitary transformation, the two-point measurement protocol will be performed on the battery. Therefore, the cumulant generating function of work defined in Eq.~\eqref{eq:workDef} is given by
\begin{align*}
\Phi(\eta,\tau) &= \ln \mathrm{Tr}\left[e^{\eta\Ham_B} {\cal U}_{S,B,1}(\tau)
e^{-\eta\Ham_B} \rho_{S,B}(0) {\cal U}_{S,B,1}^{\dagger}(\tau)\right],
\end{align*}
where $\rho_{S,B}(0) = \mathcal{G}_S(\lambda_0)\otimes \ket{x}\bra{x}_B$, and ${\cal U}_{S,B,1}(\tau)$ satisfies Eq.~\eqref{eq:ConditiononU1}. Using this property of ${\cal U}_{S,B,1}(\tau)$, we show that (see Appendix~\ref{AppA2} for details)
\begin{align}\label{mainidentity}
\Phi(\eta,\tau) &= -\frac{\eta}{\beta} S_{1-\frac{\eta}{\beta}}\left( \tilde{\rho}_S(\tau)\, ||\,\mathcal{G}_S(\lambda_\tau)\right) - \eta \Delta F,
\end{align}
where $\tilde{\rho}_S(\tau)$ is defined as
\begin{equation}\label{newrhoS}
\tilde{\rho}_S(\tau) = \left(\mathrm{Tr}_B\left[  {\cal U}_{S,B}(\tau) \,\left( \rho_S(0)^{\gamma}\otimes\rho_B(0)\right)\, {\cal U}^{\dagger}_{S,B}(\tau)  \right]\right)^{1/\gamma}
\end{equation}
with $\gamma := 1-\eta/\beta$.
A comparison between Eq.~\eqref{mainidentity} with Eq.~\eqref{eq:central} shows the mutual similarity, the only difference being in the first argument of the quantum R\'enyi-$\alpha$ divergence, namely in the $\tilde{\rho}_S(\tau)$ in place of $\rho_S(\tau)$. The former in fact now depends on the state of the newly introduced battery and on the operation $\mathcal{U}_{S,B}(\tau)$ performed on the $S,B$ compound and keeps track of the fact that the energy statistics is measured and reconstructed on the battery rather than on the system. If one substitutes $\gamma=1$, one recovers Eq.~\eqref{eq:equality_Sirr} exactly. However, the generic $\gamma$ dependence implies that the second order correction terms might take on a more complicated form, when $\rho_{S,B}(\tau)$ contains correlations. It is interesting to note that a similar observation has been made in Ref.\ \cite{alhambra2016fluctuating} (Section IV, Eq.~(30)), where higher order moments of the work distribution could not be directly analyzed due to correlating terms between system and battery energy.

In order to further compare this result with that of the resource theory setting, let us assume that the final joint state $\rho_{S,B}(\tau)$ in Eq.~\eqref{eq:redCPTP} is a product state $\rho_S(\tau)\otimes\rho_B(\tau)$. This implies that $\tilde{\rho}_S(\tau) =\rho_S(\tau)$, and thus one recovers the identity in Eq.~\eqref{eq:central}, the set of lower bounds on $\langle S_{\text{irr}} \rangle$ in Eq.~\eqref{eq:mainres1}, and also the second moment of work distribution given by Eq.~\eqref{varW}. 

{\it Comparison of the two approaches.}
In Ref.\ \cite{brandao2015second}, a family of generalized second laws has been derived in the resource theory setting. These laws form a set of necessary and sufficient conditions for single-shot state transformations, on a single-copy of $\rho_S(0)\rightarrow\rho_S(\tau)$ via thermal operations. Furthermore, one may utilize these conditions to calculate the amount of deterministic work required for this process. This is modelled by requiring that the state transition $\rho_S(0)\otimes \ket{E_0}\bra{E_0}_B\rightarrow\rho_S(\tau)\otimes \ket{E_\tau}\bra{E_\tau}_B$ satisfies all the generalized second laws, and the amount of work invested is given by $W_{\rm det} = E_0-E_\tau$. Applying the generalized second laws to this scenario tells us that the amount of deterministic work used in bringing the system from $\rho_S(0)$ to $\rho_S(\tau) $ is (\cite{brandao2015second}, Appendix I)
\begin{equation}\label{eq:ResTheory}
W_{\rm det} \geq\mathcal{F}_{\alpha}\left(\rho_S(\tau),\mathcal{G}_S(\lambda_\tau)\right) - \mathcal{F}_{\alpha}\left(\rho_S(0),\mathcal{G}_S(\lambda_0)\right) ,
\end{equation}
for all $\alpha\geq 0$, with
\begin{equation}\label{Falpha}
\mathcal{F}_{\alpha}\left(\rho(t),\mathcal{G}(\lambda_t)\right) := -\beta^{-1} \left[ \ln Z(t) - S_{\alpha}\left(\rho(t)\, ||\, \mathcal{G}_S(t)\right) \right].
\end{equation}
If the initial state is assumed to thermal i.e., $\rho_S(0) = \mathcal{G}_S(\lambda_0) $, then $\mathcal{F}_{\alpha}\left(\rho_S(0),\mathcal{G}_S(\lambda_0)\right) = -\beta^{-1}\ln Z_S(\lambda_0) $. Therefore, if one defines the quantity $S_{\rm irr}^{\rm det}:=\beta (W_{\rm det}-\Delta F)$, then  Eq.~\eqref{eq:ResTheory} reads
as follows.
\begin{observation}[Resource-theoretic irreversible entropy] The irreversible entropy in a resource-theoretic
approach is lower bounded by
\begin{equation}\label{eq:main3}
S_{\text{irr}}^{\rm det}\geq S_{\alpha}\left(\rho_S(\tau)\, ||\, \mathcal{G}_S(\lambda_\tau)\right),\qquad \alpha\geq 0.
\end{equation}
\end{observation}

Comparing Eq.~\eqref{eq:main3} and \eqref{eq:mainres1},
we observe a direct connection for $\eta \in (0,\beta)$, corresponding to the range $\alpha \in (0,1)$. In this parameter regime, we see that $\langle S_{\text{irr}} \rangle$ and $S_{\rm irr}^{\rm det}$ are bounded identically. Therefore, the stochastic approach sheds some light on the significance of these $\alpha$-free energies, due to their relation with the physically accessible quantity $\Phi(\eta,\tau)$~ \cite{batalhao2014experimental,an2015experimental}. 
In contrast, the qualitative difference between $S_{\rm irr}$ and $S_{\rm irr}^{\rm det}$ is captured by the regime of $\eta <0$ (corresponding to $\alpha >1$). In this regime, we have that $ S\langle S_{\text{irr}} \rangle \leq S_{\alpha}\left(\rho_S(\tau)\, ||\, \mathcal{G}_S(\lambda_\tau)\right) $ \cite{guarnieri2017full}, while for $S_{\rm irr}^{\rm det}$, Eq.~\eqref{eq:main3} still holds. This difference is largely due to the fact that Eq.~\eqref{eq:mainres1} deals with mean work, thus considered as a fluctuating quantity, while Eq.~\eqref{eq:main3} bounds the deterministic work $W_{\rm det}$. 
The best estimate for $S_{\rm irr}^{\rm det}$ is given by the $\infty-$R\'{e}nyi divergence instead, and in general $S_{\infty} \geq S_1$. 

{\it A second reconcilation.}
A second reconciliation point between the stochastic approach and the resource theory approach can be reached 
when one compares the quantity $W_n^\varepsilon$ in Eq.~\eqref{prob} to the $\varepsilon$-deterministic work of formation $W_{\mathrm{F},n}^\varepsilon$ derived in Ref.\ \cite{chubb2017beyond}. The analysis of $W_{\mathrm{F},n}^\varepsilon$ adapts also a resource theoretic approach, namely it considers work for a single-shot process.
 However, this process may be a global operation occuring on $ n $ copies of identical systems, for finite but large $ n $. In particular, one considers the amount of work $W_{\mathrm{F},n}^{\varepsilon}$ required in order to prepare $n$ identical copies of some final target state, $\rho_S^{\otimes n}$, with fidelity at least $1-\varepsilon$. The quantity of interest $W_{\mathrm{F},n}^{\varepsilon}$ is defined as follows: For some fixed energy value $E$, and some parameter $n$, consider the minimum integer $m$ such that the following transition on $m+n$ systems with identical, time-independent Hamiltonians,
\begin{equation}
\ket{E}\bra{E}^{\otimes m} \otimes \mathcal{G}^{\otimes n}\rightarrow \sigma, ~\quad \hat F(\sigma, \mathcal{G}^{\otimes m}\otimes \rho^{\otimes n})\geq 1-\varepsilon,
\end{equation}
is possible via thermal operations with a bath at inverse temperature $\beta$, $\mathcal{G}$ being the Gibbs state, and 
$\hat F(\rho,\sigma)$ denoting Uhlmann's fidelity \cite{uhlmann1976transition}. The amount of work is given by $W_{\mathrm{F},n}^{\varepsilon}=mE$. 
In Ref.\ \cite{chubb2017beyond}, it is shown that 
\begin{equation}
W_{\mathrm{F},n}^{\varepsilon}\approx \beta^{-1} \left[n D(\rho\|\mathcal{G})+\sqrt{n V(\rho\|\mathcal{G})} f(\varepsilon)\right],
\end{equation}
where $f(\varepsilon) > 0$. 
Comparing the expressions for $W_n^\varepsilon$ captured in Observation 3 (taking into account that when the initial and final Hamiltonian coincide, $\Delta F=0$) and $W_{{\rm F},n}^\varepsilon$, which were defined using very different approaches, we see that nevertheless they are in qualitative agreement with one another.

{\it Outlook.} In this work, we have brought two approaches to quantum thermodynamics significantly
 closer  to each other. While the approaches taken are radically different in mindset, they give rise to 
 expressions formally providing similar or identical predictions, specifically when this line of thought is
 applied to  notions of work extraction in quantum thermodynamics. It is the hope that this reconciling
 work can significantly contribute to the emerging theory of quantum thermodynamics.

{\it Acknowledgements.} 
We thank M. Campisi for valuable discussions and feedback.
G.~G. acknowledges the support of the Czech Science Foundation (GACR)
(grant No. GB14-36681G). J.~G. is supported by a SFI Royal Society University Research Fellowship. M. P. is supported by the DfE-SFI Investigator Programme (grant 15/IA/2864), the H2020 Collaborative Project TEQ (grant 766900), and the Royal Society. J.~E. acknowledges funding of the ERC (TAQ), the DFG (EI 519/7-1, EI 519/14-1, CRC 183), N.~H.~Y.~N. from the Alexander von Humboldt Foundation. 

\bibliographystyle{apsrev4-1}


%

\newpage
\onecolumngrid
\appendix

\section{Details of calculations} 
\label{AppA}

\subsection{Proof of Eq.~\eqref{eq:central}}

In the following we will explicitly derive the identity in Eq.~\eqref{eq:central}, which connects the cumulant generating function of work statistics to an opportune $\alpha-$R\'{e}nyi divergence. Here, we make use of the fact that the initial state is given by
\begin{align}
\rho_S(0) = \mathcal{G}_S(\lambda_0) = \frac{1}{Z(\lambda_0)} e^{-\beta H(\lambda_0)}. 
\end{align}
For all values of $\eta \in (-\infty,\infty) \backslash \lbrace 0,\beta \rbrace$, the cumulant generating function then reads as
\begin{align}\label{eq:2}
\Phi(\eta,\tau) &= \ln\mathrm{Tr}_{S}\left[e^{-(\eta/2)\Ham(\lambda_\tau)} U(\tau)
e^{(\eta/2)\Ham(\lambda_0)} \rho_S(0) e^{(\eta/2)\Ham(\lambda_0)} U^{\dagger}(\tau) e^{-(\eta/2)\Ham(\lambda_\tau)}\right]\notag\\
&=  \ln\mathrm{Tr}_{S}\left[e^{-\eta \Ham(\lambda_\tau)} U(t)
e^{\eta\Ham(\lambda_0)} \frac{e^{-\beta\Ham(\lambda_0)}}{Z(\lambda_0)} U^{\dagger}(\tau) \right]\notag\\
&=  \ln\mathrm{Tr}_{S}\left[e^{-\eta \Ham(\lambda_\tau)} U(\tau)
 \frac{e^{-(\beta-\eta) \Ham(\lambda_0)}}{Z(\lambda_0)} U^{\dagger}(\tau) \right] \notag\\
&=  \ln\mathrm{Tr}_{S}\left[\left(\frac{e^{-\beta\Ham(\lambda_\tau)}}{Z(\lambda_\tau)}\right)^{\eta/\beta} U(\tau) \,\left(\frac{e^{-\beta\Ham(\lambda_0)}}{Z(\lambda_0)}\right)^{1-\eta/\beta} U^{\dagger}(\tau) \right] + \ln \left[\frac{Z(\lambda_\tau)}{Z(\lambda_0)}\right]^{\eta/\beta}\notag\\
&=  \ln\mathrm{Tr}_{S}\left[\left( \mathcal{G}_S(\lambda_\tau)\right) ^{\eta/\beta}\, \left(\rho_S(\tau)\right)^{1-\eta/\beta}\right] + \ln \left[\frac{Z(\lambda_\tau)}{Z(\lambda_0)}\right]^{\eta/\beta}\notag\\
&=  \left(\frac{\eta}{\beta}-1\right) S_{\frac{\eta}{\beta}}\left(\mathcal{G}_S(\lambda_\tau)\, ||\,\rho_S(\tau)\right) + \frac{\eta}{\beta} \ln \frac{Z(\lambda_\tau)}{Z(\lambda_0)} \notag\\
&= -\frac{\eta}{\beta}S_{1-\frac{\eta}{\beta}}\left(\rho_S(\tau)\, ||\,\mathcal{G}_S(\lambda_\tau)\right) - \eta \Delta F.
\end{align}

In the last line, a skew-symmetry property of the $\alpha-$ R\'enyi divergence has been used, namely that for $ \forall \alpha \neq 0,1$, we have
\begin{equation}\label{eq:skewsym}
S_{\alpha}(\rho||\sigma) = \frac{\alpha}{1-\alpha} S_{1-\alpha}(\sigma||\rho).
\end{equation}
Alternatively, one can also achieve this by applying the cyclic property of the trace operation in the third last line, i.e. by using the fact that $\tr_S(AB)= \tr_S(BA)$.

\subsection{Proof of Eq.\ \eqref{mainidentity}}
\label{AppA2}

We know that if the global unitary $\mathcal{U}$ on joint systems $S,B,C$ the requirement ${\cal U}(\tau) = {\cal U}_{S,B,1}(\tau)\otimes\ket{\lambda_\tau}\bra{\lambda_0}_C+ {\cal U}_{S,B,2}(\tau)\otimes\ket{\lambda_0}\bra{\lambda_\tau}_C$, then it satisfies Eq.~\eqref{eq:globalUS,B,C} and also gives an effective unitary $U_{S,B}(\tau)=U_{S,B,1}(\tau)$ on the reduced state of the system and battery. This unitary, in particular, changes the Hamiltonian on the system from $\mathcal{H}_S(\lambda_0)$ to $\mathcal{H}_S(\lambda_\tau)$, as described in Eq.~\eqref{eq:ConditiononU1}. Moreover, note that since $\mathcal{H}_S (\lambda)$ always commutes with $\mathcal{H}_B$, we see how the two point measurement scheme acting on the battery can be directly related to the measurement statistics done on the system. In particular,
\begin{align}\label{eq:imprel1}
&\mathcal{U}_{S,B}(\tau) \left[\Ham_S(\lambda_0)+ \Ham_B \right]\mathcal{U}^{\dagger}_{S,B}(\tau) = \Ham_S(\lambda_\tau) + \Ham_B \notag,\\
\Rightarrow\qquad\qquad & \mathcal{U}_{S,B}(\tau) e^{-\eta [\Ham_B + \Ham_S(\lambda_0)]}\mathcal{U}^{\dagger}_{S,B}(\tau) = e^{-\eta[\Ham_S(\lambda_\tau) + \Ham_B ]}\notag,\\
\Rightarrow\qquad\qquad & \mathcal{U}_{S,B}(\tau) e^{-\eta\Ham_B}\mathcal{U}^{\dagger}_{S,B}(\tau) \mathcal{U}_{S,B}(\tau)  e^{-\eta\Ham_S(\lambda_0)}\mathcal{U}^{\dagger}_{S,B}(t) = e^{-\eta\Ham_S(\lambda_\tau)}e^{-\eta\Ham_B }\notag,\\
\Rightarrow\qquad\qquad & \mathcal{U}_{S,B}(\tau) e^{-\eta\Ham_B } = e^{-\eta\Ham_S(\lambda_\tau)} e^{-\eta\Ham_B} \,\mathcal{U}_{S,B}(\tau)e^{\eta\Ham_S(\lambda_0)}.
\end{align}
Let us consider the two-time measurement protocol on the battery $B$ as explained in the main text. The cumulant generating function is
\begin{align}\label{eq:TPMonBattery}
\Phi(\eta,\tau) &= \ln\mathrm{Tr}_{S,B,C}\left[e^{\eta\Ham_B} \mathcal{U}(\tau)
e^{-\eta\Ham_B} \,\rho_{S,B,C}(0)\,\mathcal{U}^{\dagger}(\tau)\right]\notag\\
&= \ln\mathrm{Tr}_{S,B}\left[e^{\eta\Ham_B} \mathcal{U}_{S,B}(\tau)
e^{-\eta\Ham_B} \,\rho_{S,B}(0)\,\mathcal{U}_{S,B}^{\dagger}(\tau)\right],
\end{align}
where we have substituted $\rho_C(0)=\Pi_{\lambda_0,C}$ and used Eq.~\eqref{eq:globalUS,B,C}, and traced out system $C$. We first note that, in order to evaluate $\Phi (\eta,\tau)$, the relation $\rho_B(0)^\alpha \propto \rho_B(0)$ is needed in our calculations, in order to simplify the cumulant generating function down to terms that only involve the system $S$. One can understand this intuitively: if the initial battery starts out with some non-trivial energy distribution, and if one defines the work statistics via TPM on the battery, this will depend not only on work fluctuations from the system but also on the prior distribution on the battery. Furthermore, $\rho_B(0)^\alpha \propto \rho_B(0)$ is satisfied if and only if 
\begin{equation}
\rho_B(0) = \frac{1}{d_R}\Pi_R 
\end{equation}
is a maximally mixed state on a subspace $R$ of dimension $d_R$. In particular, if $\rho_B$ is of such a form, then
\begin{equation}\label{eq:MMbattery}
\rho_B(0)^\alpha = d_R^{1-\alpha}
\rho_B(0).
\end{equation}
Later on, we shall see that in order to maximize the cumulant generating function, $d_R = 1$, which means that the battery initial state $ \rho_B(0)$ is a pure Gaussian state approximating a state with definite position.
Now, by evaluating Eq.~\eqref{eq:TPMonBattery} by making use of Eq.~\eqref{eq:imprel1}, we have
\begin{align}\label{eq:res1a}
\Phi(\eta,\tau) &= \ln\mathrm{Tr}_{S,B}\left[e^{\eta\Ham_B}\mathcal{U}_{S,B}(\tau)
e^{-\eta\Ham_B} \,\rho_{S,B}(0)\,\mathcal{U}_{S,B}^{\dagger}(\tau)\right] \notag\\
&= \ln\mathrm{Tr}_{S,B}\left[e^{\eta\Ham_B}  \left(e^{-\eta\Ham_S(\lambda_\tau)}e^{-\eta\Ham_B} \mathcal{U}_{S,B}(\tau)e^{\eta\Ham_S(\lambda_0)}\right) \rho_{S,B}(0)\,\mathcal{U}_{S,B}^{\dagger}(\tau)\right] \notag\\
&= \ln\mathrm{Tr}_{S,B}\left[e^{-\eta\Ham_S(\lambda_\tau)}\mathcal{U}_{S,B}(t)
e^{\eta\Ham_S(\lambda_0)} \, \left[\mathcal{G}_S(\lambda_0) \otimes\rho_B(0)\right] \,\mathcal{U}_{S,B}^{\dagger}(t)\right]\notag\\ 
&=  \ln\mathrm{Tr}_{S,B}\left[ \left(\frac{e^{-\beta\Ham_S(\lambda_\tau)}}{Z(\lambda_\tau)}\right)^{\eta/\beta}\!\!\!\!\otimes\Id_B \, \mathcal{U}_{S,B}(t) \,\left(\frac{e^{-\beta\Ham_S(\lambda_0)}}{Z(\lambda_0)}\right)^{1-\eta/\beta}\!\!\!\!\!\!\!\! \otimes\rho_B(0) \,\mathcal{U}_{S,B}^{\dagger}(t)\right] + \frac{\eta}{\beta}\ln\frac{Z(\lambda_\tau)}{Z(\lambda_0)}\notag\\ 
&=  \ln\mathrm{Tr}_{S,B}\left[ \left(\mathcal{G}_S(\lambda_\tau)\otimes\Id_B\right)^{\eta/\beta} \, \mathcal{U}_{S,B}(t) \,\left(\mathcal{G}_S(\lambda_0)\otimes \rho_B(0)\right)^{1-\eta/\beta}\,\mathcal{U}_{S,B}^{\dagger}(t)\right] + \frac{\eta}{\beta}\ln\frac{Z(\lambda_\tau)}{Z(\lambda_0)} -\frac{\eta}{\beta}\ln d_R.
\end{align}
Besides other manipulations which are straightforward, the last line is obtained by assuming Eq.~\eqref{eq:MMbattery}. Observing Eq.~\eqref{eq:res1a}, since the last term is always non-positive for $\eta >0$, in order to maximize it we take $d_R = 1$, and this term vanishes. We also observe that within the first term, we have an expression ${\cal U}_{S,B}(t) \rho_{S,B}(0)^{1-\eta/\beta}{\cal U}_{S,B}(t)^\dagger$. To simplify this, note that the final state on $S,B$ is given by
\begin{equation}
\rho_{S,B}(t) = {\cal U}_{S,B}(t) \rho_{S,B}(0){\cal U}_{S,B}(t)^\dagger = \sum_i p_i \ket{f_i}\bra{f_i}_{S,B},
\end{equation}
where $\rho_{S,B}(0)$ has the diagonal form $\rho_{S,B}(0)=\sum_i p_i \ket{e_i}\bra{e_i}$, and $\mathcal{U}_{S,B}$ performs a unitary transformation from the ordered basis $\lbrace \ket{e_i}_{S,B}\rbrace$ to $\lbrace \ket{f_i}_{S,B}\rbrace$. In other words, for all $ i $, $ \mathcal{U}_{S,B}\ket{e_i}=\ket{f_i} $. This, together with the fact that for any $ \alpha \in\mathbb{R}$, $\rho_{S,B}(0)^\alpha = \sum_i p_i^\alpha \ket{e_i}\bra{e_i}$ gives us that
\begin{equation}\label{key}
{\cal U}_{S,B}(\tau) \rho_{S,B}(0)^{1-\eta/\beta}{\cal U}_{S,B}(\tau)^\dagger = \rho_{S,B}(\tau)^{1-\eta/\beta},
\end{equation}
which we may now substitute back into Eq.~\eqref{eq:res1a}. Finally, we have
\begin{align}\label{eq:res1}
\Phi(\eta,\tau)&= \ln\mathrm{Tr}_{S,B}\left[ \left(\mathcal{G}_S(\lambda_\tau)\otimes\Id_B\right)^{\eta/\beta} \, \left(\rho_{S,B}(\tau)\right)^{1-\eta/\beta}\right] + \frac{\eta}{\beta}\ln\frac{Z(\lambda_\tau)}{Z(\lambda_0)}\notag\\
&= \ln\mathrm{Tr}_{S}\left[ \left(\mathcal{G}_S(\lambda_\tau)\right)^{\eta/\beta} \, \mathrm{Tr}_B\left[\left(\rho_{S,B}(\tau)\right)^{1-\eta/\beta}\right]\right] + \frac{\eta}{\beta}\ln\frac{Z(\lambda_\tau)}{Z(\lambda_0)}\notag\\
&= \left(\frac{\eta}{\beta}-1\right) S_{\frac{\eta}{\beta}}\left(\mathcal{G}_S(\lambda_\tau) \,||\,\tilde{\rho}_S(\tau)\right) + \frac{\eta}{\beta} \ln \frac{Z(\lambda_\tau)}{Z(\lambda_0)}\notag\\
&= -\frac{\eta}{\beta} S_{1-\frac{\eta}{\beta}}\left( \tilde{\rho}_S(\tau)\, ||\,\mathcal{G}_S(\lambda_\tau)\right) + \frac{\eta}{\beta} \ln \frac{Z(\lambda_\tau)}{Z(\lambda_0)},
\end{align}
where the idempotency of $\rho_B(0)$ and of $\Id_B$ have been used (an operator $A$ is idempotent iff for any $\gamma\neq 0$, we have $A^\gamma = A$). Furthermore, we have defined the quantity
\begin{equation}\label{newrhoSdef}
\tilde{\rho}_S(\tau) = \left(\mathrm{Tr}_B\left[  {\cal U}_{S,B}(\tau) \,\left( \rho_S(0)^{\gamma}\otimes\rho_B(0)\right)\, {\cal U}^{\dagger}_{S,B}(\tau)  \right]\right)^{1/\gamma},\qquad \gamma := 1-\eta/\beta.
\end{equation}
Similarly, Eq.~\eqref{eq:skewsym} has been used to obtain the last line in Eq.~\eqref{eq:res1}.

\section{Details of the consistency between the family of lower bounds Eq.\ \eqref{eq:mainres1} and \cite{brandao2015second}}
\label{AppB}

According to Ref.\ \cite{brandao2015second}, a transition from a generic state $\rho$ to another state $\rho'$ is possible if and only if the family of generalized second laws applies, namely \textit{iff}
\begin{equation}\label{secondlaws}
 \mathcal{F}_{\alpha}\left(\rho,\mathcal{G}\right) \geq  \mathcal{F}_{\alpha}\left(\rho',\mathcal{G}'\right) ,\qquad \forall \alpha\geq 0,
\end{equation}
where $\mathcal{F}_{\alpha}$ is defined in Eq.\ \eqref{Falpha} and $\mathcal{G}$ denotes the Gibbs state of the system.
According to Ref.\ \cite{brandao2015second}, we assume to start an initially uncorrelated state
\begin{equation}
\rho_{S,B}(0) = \rho_S(0) \otimes \rho_B(0),
\end{equation} 
where, in line with our choice throughout the main text, we take the initial state of the battery to be $\rho_B(0)=\ket{E_0}\bra{E_0}_B$, corresponding to a pure energy eigenstate with energy $E_B(0)$. Moreover, in the weak coupling regime assumed in Ref.\ \cite{brandao2015second} (see their Appendix I for details) implies
\begin{equation}
\rho_{S,B}(\tau) := \rho' = \rho_S(\tau) \otimes \rho_B(\tau).
\end{equation}
In what follows, $\mathcal{G}_{S(B)} $ denotes the Gibbs state of the system (battery). Furthermore, for deterministic work, the final state of $ \rho_B(\tau) = \ket{E_B(\tau)}\bra{E_B(\tau)}_B$ is a pure energy eigenstate as well.
Using the definition of $\mathcal{F}_{\alpha}$ as given by Eq.\ \eqref{Falpha}, and noting that $\mathcal{F}_{\alpha}$ is additive under tensor product, we have therefore that Eq.\ \eqref{secondlaws} reduces to
\begin{align*}
&\mathcal{F}_{\alpha}\left(\rho_S(0)\otimes\rho_B(0),\mathcal{G}_S\otimes\mathcal{G}_B\right) \geq \mathcal{F}_{\alpha}\left(\rho_S(\tau)\otimes\rho_B(\tau),\mathcal{G}'_S\otimes\mathcal{G}'_B\right) \\
&\mathcal{F}_{\alpha}\left(\rho_B(0),\mathcal{G}_B\right) -  \mathcal{F}_{\alpha}\left(\rho_B(\tau),\mathcal{G}'_B\right)\geq   \mathcal{F}_{\alpha}\left(\rho_S(\tau),\mathcal{G}'_S\right) -\mathcal{F}_{\alpha}\left(\rho_S(0),\mathcal{G}_S\right) \\
&\beta^{-1} \left[S_{\alpha}(\rho_B(0) || \mathcal{G}_B) - \ln Z_B(0) -S_{\alpha}(\rho_B(\tau) || \mathcal{G}'_B) +\ln Z_B(\tau)\right]\geq   \mathcal{F}_{\alpha}\left(\rho_S(\tau),\mathcal{G}'_S\right) -\mathcal{F}_{\alpha}\left(\rho_S(0),\mathcal{G}_S\right).
\end{align*}
To further simplify these expressions, let us note that $S_{\alpha}(\rho_B(0) || \mathcal{G}_B)$ can be further evaluated. Since $ \rho_B(0) $ is diagonal in the energy eigenbasis by construction, $\rho_B(0)$ and $\mathcal{G}_B$ also commute. Their $\alpha-$R\'{e}nyi divergence is simply given by
\begin{align*}
S_{\alpha}(\rho_B(0) || \mathcal{G}_B) &= \frac{1}{\alpha-1} \ln \tr \left[ (\ket{E_B(0)}\bra{E_B(0)})^{\alpha} \left(\frac{e^{-\beta \mathcal{H}_B}}{Z_B(0)}\right)^{1-\alpha} \right] = \frac{1}{\alpha-1} \ln\left(\frac{e^{-\beta E_B(0)}}{Z_B(0)}\right)^{1-\alpha} = \beta E_B(0) + \ln Z_B(0).
\end{align*}
This expression holds for a pure battery state, and since one is concerned with deterministic work here, a similar expression holds for $ S_{\alpha}(\rho_B(\tau) || \mathcal{G}_B)$ as well. Substituting both expressions into the previous result, we end up with
\begin{align*}
E_B(0) - E_B(\tau) &\geq \mathcal{F}_{\alpha}\left(\rho_S(\tau),\mathcal{G}'_S\right)-\mathcal{F}_{\alpha}\left(\rho_S(0),\mathcal{G}_S\right),
\end{align*}
from which, by simply using the definition of work given by the energy difference in the battery $ W = E_B(0) - E_B(\tau) $, we obtain Eq.\ \eqref{eq:ResTheory}. 

\end{document}